\documentclass[12pt]{article}
\usepackage[cp866]{inputenc}
\usepackage[english]{babel}

\usepackage{latexsym}
\usepackage{amsmath}
\usepackage{amssymb}
\usepackage{indentfirst}
\usepackage{color}
\usepackage{graphicx}
\usepackage{bm}
\usepackage{maybemath}
\usepackage{revsymb4-1}

 \newcommand{\be}{\begin{equation}}
\newcommand{\ee}{\end{equation}}
\newcommand{\dst}{\displaystyle}
\newcommand{\fr}[2]{\frac{{\dst #1}}{{\dst #2}}}

\newcommand{\bea}{\begin{eqnarray}}
\newcommand{\eea}{\end{eqnarray}}

\newcommand{\dd}{\mathrm{d}}
\newcommand{\EE}{\mathrm{e}}

\title{Bound--free pair production in relativistic nuclear collisions from the NICA to the HE LHC colliders}
\author{D.\,A. Bauer$^1$, D.\,V.~Karlovets$^2$, V.\,G.~Serbo$^{1,3}$}

\date{\small{$^1$ Novosibirsk State University, RUS-630090, Novosibirsk, Russia,\\
$^2$ Tomsk State University, Lenina Ave.\,36, 634050 Tomsk, Russia,\\
$^3$ Sobolev Institute of Mathematics, RUS-630090, Novosibirsk, Russia}}

\begin{document}
\maketitle

\begin{abstract}
We consider the electron-positron pair production in relativistic heavy ion collisions,
in which the produced electron is captured by one of the nuclei resulting, thus,
in the formation of a hydrogen--like ion. These ions emerge from the collision point and hit the vacuum chamber wall inside superconducting magnets.
Therefore, this process may be important for the problems of beam life time and for the quenching the irradiated magnet.
A theoretical investigation for such a bound-free pair production (BFPP) at the colliders from NICA to HE LHC is presented.
We obtain an approximate universal formula for the total cross section of the process. We compare it with the results of available numerical calculations
and estimate that an accuracy of our calculations is better than $30$ \% at the energies of the NICA collider
and becomes of the order of a few percent for the RHIC and HE LHC colliders.
Based on the obtained results, the detailed calculations are performed for future experiments at the NICA collider.
We find that the expected BFPP cross sections for the Au$^{79+}$--Au$^{79+}$ and Bi$^{83+}$--Bi$^{83+}$ collisions
are in the range from $10$ to $70$ barn, while for the p--Au$^{79+}$ and {p--Bi$^{83+}$} collisions they are in the range of a few mbarn.

\end{abstract}

\section{Introduction}

In this paper we discuss the process of the bound--free pair production,
in which the produced electron is captured by one of the nuclei resulting, thus, in the formation of a hydrogen--like ion,
\begin{equation}
\label{eq_process}
Z_1 + Z_2 \to Z_1 + e^+ + (Z_2+e^-)_{1s_{1/2}}  \,,
\end{equation}
where the bound system is denoted by the round brackets. For definiteness, we assume (if it is not stated otherwise)
that the electron is captured by the second nucleus into the ground ionic state.

One of the dominant processes at the colliders with heavy ions is electron--positron pair production, whose cross section is huge.
For example, the cross section of the creation of a \textit{free} $e^+ e^-$ pair at the Pb$^{82+}$--Pb$^{82+}$ collisions on LHC
reaches hundreds of kilobarns \cite{BuG75,IvS99,LeM02,LeM09,Bau07}. In contrast, the BFPP process has orders of magnitude
smaller cross section than free-free pair production. Nevertheless, its investigation is of great importance for two reasons.
In free-free $e^+e^-$ production, the scattered nuclei lose only a very small fraction of their energy and acquire tiny scattering angles,
and thus do not leave the beam. In contrast, if one of the colliding ions captures an electron, it changes its charge and
it is bent out from the beam in a collider magnet. The corresponding cross section is usually larger that the total hadronic cross section.
Therefore, the reaction~\eqref{eq_process} may be one of the important processes which limits the luminosity of colliders.
Besides, the secondary beams of down-charged ions emerging from the collision point hit a beam-pipe and deposit
a considerable portion of energy at a small spot, which may in turn lead to the quenching of superconducting magnets~\cite{Bau02,JoB05,Bru07,BrB09}.
That is why this process has a considerable interest both from experimental and theoretical point of view.
In particular, one can point out the old experiment at the CERN SPS~\cite{SPS} and the recent one at the ALICE (LNC)~\cite{ALICE-2019},
as well as the theoretical papers~\cite{Beccker1987,MHHTB-2001} (see also literature therein).
Many aspects in calculations of the total and differential cross sections of the BFPP process have been
considered in detail in paper~\cite{AJSS-2012} oriented to application at the LHC colliders.

One of the motivation for the present study was the request to evaluate the significance of this process for the NICA collider under construction.
At this collider, collisions of the gold and bismuth nuclei are foreseen with the total energy in the range of $\sqrt{s_{NN}}=4\div 11$ GeV/nucleon,
as well as the proton-nucleus collisions with the proton energy up to 12 GeV -- see the description of the project in paper~\cite{NICA}
and the present status in~\cite{Conference-2020}. Unfortunately, the calculations from paper~\cite{AJSS-2012} were not applicable for NICA due to the considerably lower energy of the colliding nuclei.

In order to provide very simple but reliable estimates for future experiments with the different types of nuclei and different energies, we employ the equivalent photon approximation together with the modified Sauter approximation. Basically we follow an approach of the paper~\cite{AJSS-2012} pointing out the modifications related to lower but relativistic energies of the colliding nuclei.

In Sec.2 we remind the kinematics of the BFPP process, show how the corresponding cross section is related to that
of the photo-production process within the equivalent photon approximation, and compare the approximate cross section for collisions of $\text{Au}^{79+}$
with the exact result from Ref.~\cite{AgS97}. In Sec.3.1 we present results for the proton-nucleus and nucleus-nucleus collisions for the parameters of the NICA collider
and different nuclei, and in Sec.3.2 we estimate the accuracy of our results as a function of the collision energy. We conclude in Sec.3.3.

We use the relativistic units $c=1$, $\hbar=1$ and the notations and results of paper~\cite{AJSS-2012}. The initial state of the overall system is given by two bare nuclei
of the charges $Z_1$ and $Z_2$ and the masses $M_1$ and $M_2$, having 4--momenta $P_{1}$ and $P_{2}$; the electron mass is denoted as $m$.
Some detail of calculation can be found in paper~\cite{AJSS-2012}.

\section{Theoretical background}

\begin{figure}[h]
\begin{center}
     \includegraphics[width=7cm]{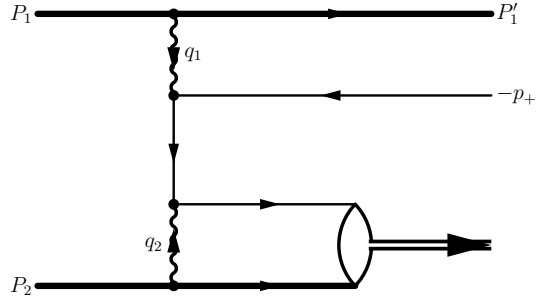}
\end{center}
\caption{Schematic Feynman diagram for the bound--free pair
production in a heavy--nucleus collision. Heavy colliding nuclei are represented by thick lines while thinner lines correspond to light fermions (electron and positron).
The trajectory of the produced electron is aligned with that of the second nucleus.}
 \label{Fig1}
\end{figure}

Our theoretical analysis of the BFPP is
based on quantum electrodynamic. In the lowest order in $Z_1\sqrt{\alpha}$ parameter, the
process (\ref{eq_process}) is described
by the diagram in Fig.~\ref{Fig1}, where the two thick lines represent
the colliding nuclei, the thinner lines correspond to
light fermions (electron and positron) and the double--line arrow just refers to the
residual hydrogen--like ion. Inspecting this diagram, we may
interpret the BFPP as being due to the
interaction of the second (lower) nucleus with the virtual (or
equivalent) photon with the energy $\omega=q_1^0$ and virtuality
$Q^2= - q_1^2$ emitted by the first nucleus.

Our approach exploits the equivalent photon approximation (EPA)
as an \textit{approximate} method for calculating a Feynman
diagram for the corresponding process and uses the fact that
the virtual photons in the diagram are close to the mass
shell---for details, see the review~\cite{BuG75}.
Thus, the theoretical analysis of the $e^+e^-$ creation in energetic
heavy--ion collisions can be traced back to the virtual process:
\begin{equation}
\label{eq_EPA_pair_production}
\gamma^* + Z_2 \to e^+ + (Z_2 + e^-)_{1s_{1/2}} \, .
\end{equation}
The total cross section $\sigma_{Z_1Z_2}$ of the process
(\ref{eq_process}) can be expressed approximately in terms of the
cross sections $\sigma_{\gamma Z_2}$ for the (real) pair photo-production (\ref{eq_EPA_pair_production})
as:
\begin{equation}
\label{eq_EPA_cross_section}
\sigma^{\rm app}_{Z_1Z_2}(\gamma_L) = \int_{2m}^{\omega_{L,\max}} {\dd}n(\omega_L)\, \sigma_{\gamma Z_2}(\omega_L) \, .
\end{equation}
In this expression, we denote by ${\dd}n$
the number of the equivalent photons and, moreover, neglect the interaction
between the emitted positron and the first nucleus. Below we will evaluate
${\dd}n$ and $\sigma_{\gamma Z_2}$ in the
rest frame of the second nucleus, which finally
constitutes the hydrogen--like ion. The energy of the first nucleus in such a frame is
 \be
 E_L=\gamma_L M_1=\fr{P_1P_2}{M_2},
 \ee
while the energy of the equivalent
photon, emitted by the first nucleus (cf. Fig.~\ref{Fig1}), reads
\begin{equation}
\label{eq_photon_energy_proj_frame}
\omega_L = \frac{q_1P_2}{M_2}\,.
\end{equation}
For the collision of identical nuclei (when $Z_1=Z_2=Z$), the Lorentz-factor $\gamma_L$ of the first nucleus in the rest frame of the second nucleus has a relation
 \be
 \gamma_L=2\gamma^2-1
 \label{gLg}
 \ee
with the Lorentz-factor $\gamma$ of a single nucleus in the centre-of-mass system.

It was found out in paper~\cite{AJSS-2012} that the pair photo--production cross section $\sigma_{\gamma Z_2}$
has a good analytical approximation in the form
 \bea
 \label{sigma-gammaZ}
\sigma^{\rm app}_{\gamma Z_2}(\omega_L)&=&f(Z_2)\,4\pi \fr{Z_2^5 \alpha^6}{m^2}\,G(\omega_L/m),
\\
G(y) &=& \fr{\sqrt{y(y-2)}}{y^4}\left[y^2-\fr 43 y + \fr 53
-\fr{y+1}{\sqrt{y(y-2)}}\,\ln\left(y-1+\sqrt{y(y-2)}\right) \right],
\label{G(y)}
 \eea
where $f(Z)$ is given in Table 1 (taken from~\cite{AgS97}) and the expression
$4\pi \left(Z_2^5 \alpha^6/m^2\right)\,G(\omega_L/m)$ is the photo-production
cross section in the lowest order in $Z\alpha$
(it can be obtained from Sauter's result for the photoelectric effect~\cite{Sau31}).
%
 \begin{center}
 \centerline{ Table 1}
 \vspace{3mm}
 \begin{tabular}{|c|c|c|c|c|c|c|c|}
  \hline
  $Z$ & 1 & 8 & 26 & 55 & 79 & 82 & 92 \\ \hline
  $f(Z)$ & 0.971 & 0.798 & 0.518 & 0.293 & 0.222 & 0.216 & 0.196\\
  \hline
\end{tabular}
\end{center}
 \vspace{5mm}
 %
\begin{figure}[h]
\begin{center}
     \includegraphics[width=12cm]{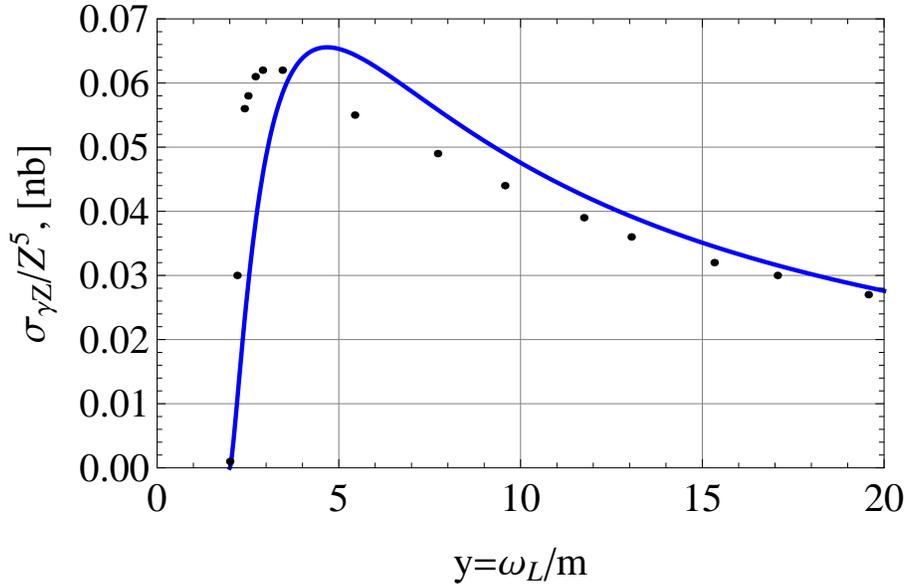}
\end{center}
\caption{The approximate cross section $\sigma^{\rm app}_{\gamma Z}/Z^5$ in nanobarn from Eq.~\eqref{sigma-gammaZ} in dependence on $y=\omega_L/m$ for the Au$^{79+}$--Au$^{79+}$ (blue solid line) and points from the exact result
of Ref.~~\cite{AgS97}}
 \label{Fig2}
\end{figure}
To prove this statement we compare in Fig.~\ref{Fig2} the approximate cross section $\sigma^{\rm app}_{\gamma Z}/Z^5$ from Eq.~\eqref{sigma-gammaZ} for the Au$^{79+}$--Au$^{79+}$ (blue solid line) and points taken from the exact result of Ref.~~\cite{AgS97}. It is clearly seen from Fig.~\ref{Fig2} that presented approximation is really
close to the exact result.

The number of equivalent photons reads (see Eq. (28) in \cite{AJSS-2012})
 \be
 {\dd}n(\omega_L)= \fr{Z_1^2 \alpha}{\pi}\,\fr{{\dd}\omega_L}{\omega_L}
 \left[\ln{\fr{\gamma^2_L Q^2_{\max}}{\omega^2_L}}-1 \right]
 \ee
with $Q^2_{\max}=4m^2$ and the value\footnote{In paper~\cite{AJSS-2012} it was considered the case of very high
energies, $\gamma \geq 100$, with the approximate relation $\gamma_L=2\gamma^2$ and the value
$\omega_{L,\max}=\infty$ which can be used in that case due to the fast convergency of the integral in equation~\eqref{eq_EPA_cross_section}.}
 \be
 \omega_{L,\max} = m y_{\max}=m \fr{2\gamma_L}{\sqrt{\EE}}=1.21\, m \gamma_L,\;\;\EE=2.718...,
 \ee
which is determined by the requirement ${\dd}n(\omega_L)/{\dd}\omega_L > 0$.

\begin{figure}[h]
\begin{center}
     \includegraphics[width=12cm]{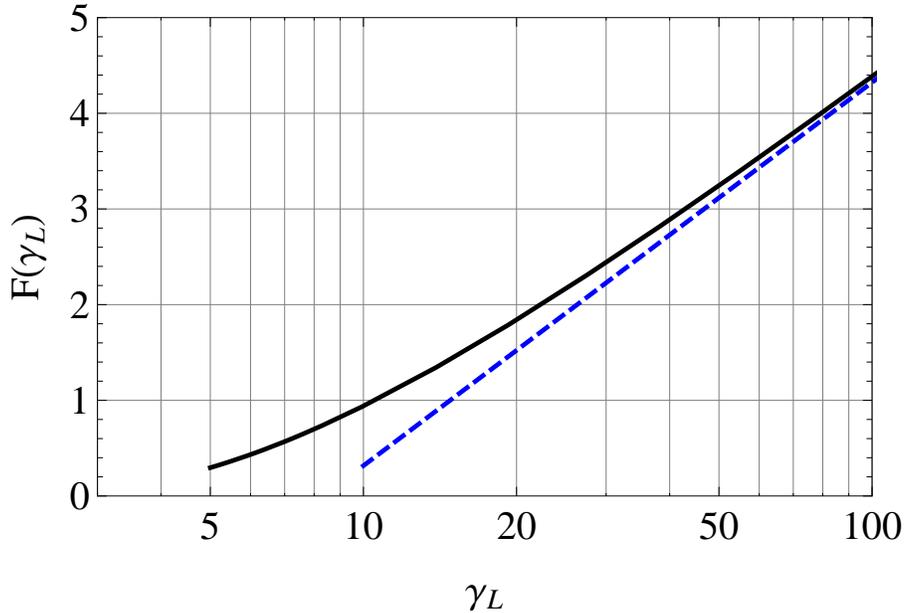}
\end{center}
\caption{The function $F(\gamma_L)$ from Eq.~\eqref{final} (black solid line) and its asymptotics $F_{\rm asymp}(\gamma_L)$ from Eq.~\eqref{Fasym} (blue dashed line)}
 \label{Fig3}
\end{figure}

As a result, we obtain the following approximate expression for the cross section
of the process
 \be
\sigma^{\rm app}_{Z_1 Z_2}(\gamma_L)=f(Z_2)\,\fr{Z_1^2 Z_2^5\alpha^7}{m^2}\,F(\gamma_L),\;\;
F(\gamma_L)=8\,\int_2^{y_{\max}}\,G(y)\,
\ln\fr{y_{\max}}{y}\,\fr{{\dd}y}{y}.
\label{final}
 \ee
The important universal function  $F(\gamma_L)$ incorporates all dependence on the collider's energy.
It  is shown in Fig~\ref{Fig3} together with its asymptotic
expression at $\gamma_L \gg 1$:
 \be
 F_{\rm asymp}(\gamma_L)= 1.74 \,\ln\gamma_L- 3.69.
 \label{Fasym}
 \ee
The latter coincides with the function $F(\gamma_L)$ with the accuracy better than $5$ \% at $\gamma_L> 45$.
Some values of function $F(\gamma_L)$ in the region of the NICA energies, $\gamma =2\div 5.5$, are given
in Table~2.
 \begin{center}
 Table 2\\
 \vspace{3mm}
\begin{tabular}{|c|c|c|c|c|c|c|c|c|}
  \hline
  $\gamma_L=(P_1P_2)/(M_1M_2)$ & 7 & 11.5 & 17 & 23.5 & 31 & 39.5 & 49 & 59.5 \\ \hline
  $\gamma=\sqrt{(\gamma_L+1)/2}$ & 2 & 2.5 & 3 & 3.5 & 4 & 4.5 & 5 & 5.5 \\ \hline
  $F(\gamma_L)$ & 0.569 & 1.11 & 1.61 & 2.07 & 2.49 & 2.87 & 3.21 & 3.53 \\
  \hline
\end{tabular}
\end{center}

 \section{Results and discussion}

 \subsection{Results for the NICA collider}

\begin{figure}[h!]
\begin{center}
     \includegraphics[width=12cm]{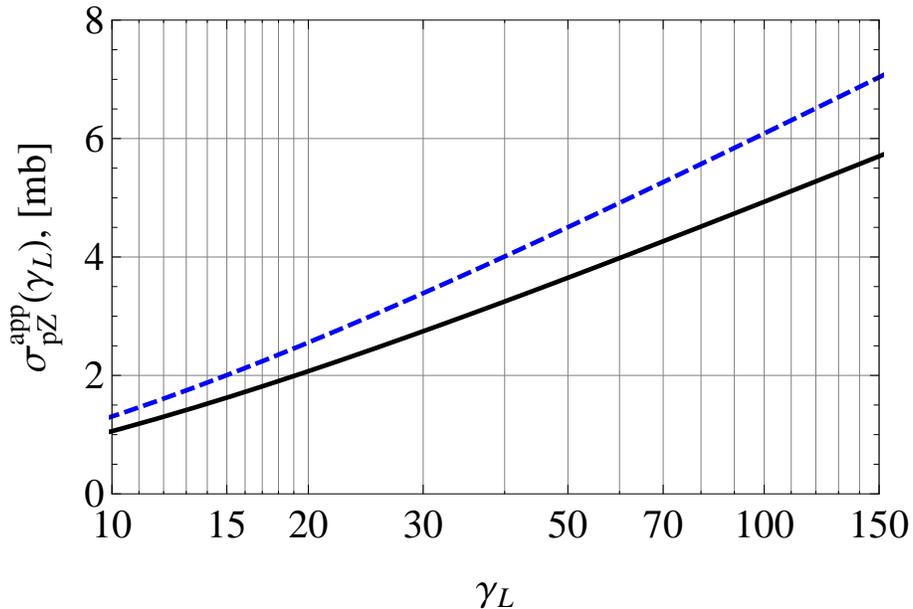}
\end{center}
\caption{Cross sections in mbarn from Eq.~\eqref{final} in dependence on $\gamma_L=(P_1P_2)/(M_pM_2)$ for
the p--Au$^{79+}$ (black solid line) and p--Bi$^{83+}$ (blue dashed line)  collisions }
 \label{Fig4}
\end{figure}

{\bf A. Proton-nucleus collisions.} The BFPP cross sections for the p--Au$^{79+}$ and p--Bi$^{83+}$ collisions are shown in Fig.~\ref{Fig4}. Here we use $F(Z=83)=0.214$ obtained by fit to values given in Table 1. In particular,
the cross section of the BFPP in the p--Bi$^{83+}$ collision at the NICA collider with the proton energy $10$ GeV and the ion energy $5$ GeV/nucleon (that is, $\gamma_L\approx 100$) is equal to $6$ mbarn, which is considerably smaller than the cross section for the hadronic interaction.

\begin{figure}[h]
\begin{center}
     \includegraphics[width=12cm]{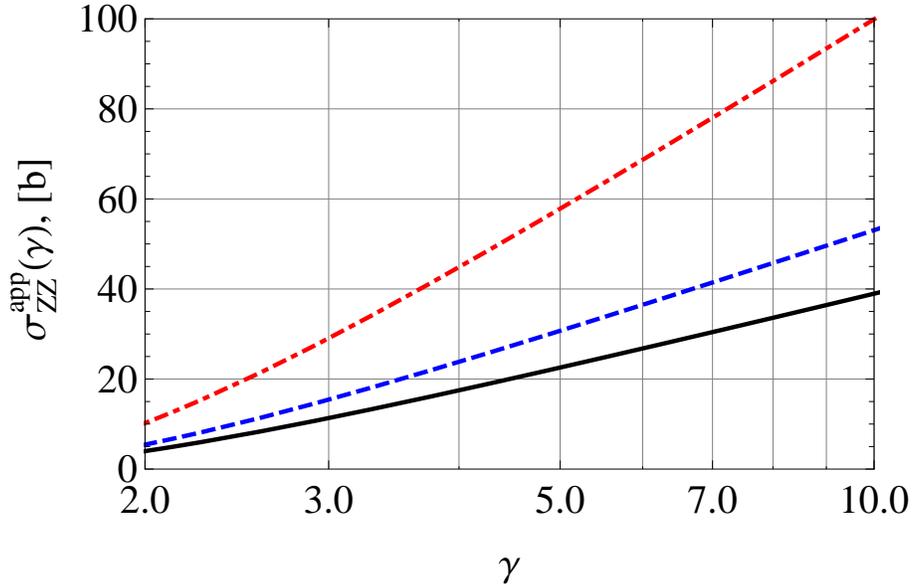}
\end{center}
\caption{Cross sections in barn from Eq.~\eqref{finalZZ} in dependence on $\gamma=\sqrt{(P_1+P_2)^2/(4M^2)}$ for
the Au$^{79+}$--Au$^{79+}$ (black solid line), Bi$^{83+}$--Bi$^{83+}$ (blue dashed line)
and U$^{92+}$--U$^{92+}$ (red dot-dashed line)  collisions }
 \label{Fig5}
\end{figure}

{\bf B. Nucleus-nucleus collisions} Now we consider the collision of the identical nuclei, $Z_1=Z_2=Z$, when
 \be
 y_{\max}=\fr{\omega_{L,\max}}{m}=\fr{2\gamma_L}{\sqrt{\EE}}=\fr{4\gamma^2-2}{\sqrt{\EE}}.
 \ee
and the corresponding BFPP cross section reads
 \be
\sigma^{\rm app}_{Z Z}(\gamma)=f(Z)\,\fr{(Z\alpha)^7}{m^2}\,F(\gamma_L=2\gamma^2-1).
\label{finalZZ}
 \ee
The cross sections for the Au$^{79+}$--Au$^{79+}$, Bi$^{83+}$--Bi$^{83+}$ and U$^{92+}$--U$^{92+}$ collisions are
shown in Fig.~\ref{Fig5}.

\subsection{Accuracy of the used approximation}

\begin{figure}[h!]
\begin{center}
     \includegraphics[width=12cm]{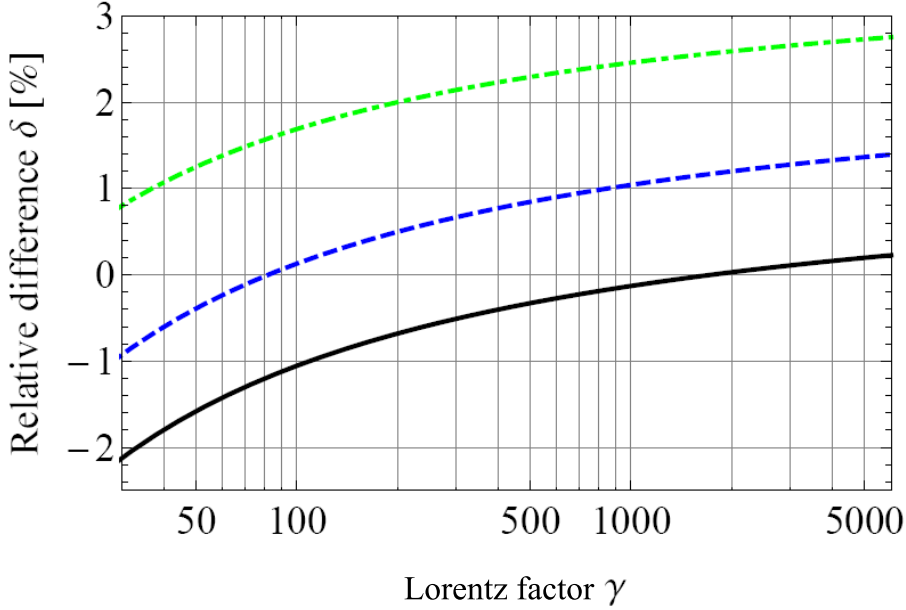}
\end{center}
\caption{The relative difference (\ref{DeltaMeier})
between rigorous relativistic results and our predictions \eqref{finalZZ} for the p--p
(green dot-dashed line) as well as  Au$^{79+}$--Au$^{79+}$ (blue dashed
line) and Pb$^{82+}$--Pb$^{82+}$ (black solid line) collisions}
 \label{Fig6}
\end{figure}

In the region of the large energy, $\gamma > 30$, there is a fine agreement of the presented
approximate calculation~\eqref{finalZZ} with the previous ones~\cite{AJSS-2012}
and with the exact calculations of Meier et al.~\cite{MHHTB-2001}. The corresponding relative difference
 \be
 \delta=\fr{\sigma^{\rm app}_{ZZ} - \sigma^{\rm Meier}_{ZZ}}{\sigma_{ZZ}^{\rm Meier}}
 \label{DeltaMeier}
 \ee
is shown in Fig.~\ref{Fig6}. It is seen that in the interval $\gamma = 30 \div 6000$ the accuracy is
better than ${3\,\%}$.

In the region of the NICA energies, $\gamma =2\div 5.5$, the accuracy is noticeably worse. That  can be seen from the comparison between our result~\eqref{finalZZ} and that of Becker at al.~\cite{Beccker1987}, where there are data for the special case of Au$^{79+}$--Au$^{79+}$ collisions with $\gamma_L = 1\div 100$. The corresponding numbers are given in Table~3, in which the last line represents the relative error of our approximation:
 \be
 \delta= \fr{\sigma^{\rm app}_{ZZ} (\ref{finalZZ})- \sigma^{\rm Becker}_{ZZ}}{\sigma_{ZZ}^{\rm Becker}}.
 \label{DeltaBecker}
 \ee
It is seen from Table 3 that $|\delta|\leq 0.3$ in the NICA interval $\gamma = 2\div 5.5$.
 \vspace{5mm}
 \begin{center}
  Table 3: Comparison of the cross sections for the case of
Au$^{79+}$--Au$^{79+}$ collisions\\
 \vspace{3mm}
 \begin{tabular}{|c|c|c|c|c|c|c|c|c|c|}
   \hline
   $\gamma=\sqrt{(\gamma_L+1)/2}$ & 2 & 2.5 & 3 & 3.5 & 4 & 4.5 & 5 & 5.5 \\ \hline
    $\sigma^{\rm app}_{Z Z}$ [b] from Eq.~\eqref{final}  & 4.0 & 7.76 & 11.3 & 14.6 & 17.5 & 20.1 & 22.5 & 24.7  \\ \hline
   $\delta$ \% & -25 & -5.4 & 3 & 12 & 17 & 26  & 25  & 30\\
   \hline
 \end{tabular}
\end{center}
 \vspace{5mm}

Another comparison can be made by the example of collisions of Pb nuclei ( $Z_1=82$) and the Au target nuclei ( $Z_2=79$ ) with the capture of the produced electron by the first nucleus. This cross section can be obtained from Eq.~\eqref{final} by exchange $Z_1\leftrightarrow Z_2$. It gives $\sigma^{\rm app}_{Z Z}= 43$ barn for $\gamma_L=168$. This number can be compared with $45$ barn from the exact calculations of Meier et al.~\cite{MHHTB-2001} and with
$44.3$ barn from the experimental result of the CERN SPS~\cite{SPS}.

\subsection{Discussion}

The simple approximate expression for the cross section at the collisions of the identical
nuclei, $Z_1=Z_2=Z$, in the range of very high-energy $\gamma = 100 \div 3000$ has been obtained in
paper~\cite{AJSS-2012}:
 \be
\sigma^{\rm app}_{Z}(\gamma)=f(Z)\,\fr{(Z\alpha)^7}{m^2}\,(3.479\,\ln\gamma-2.49).
\label{final-2012}
 \ee
Unfortunately, this expression does not work for the NICA collider with $\gamma=2\div 5.5$. For example, it leads to
the negative cross section at $\gamma=2$.
In the present paper we consider the more general case, including collisions of the non-identical
nuclei, $Z_1 \neq Z_2$, as well as the lower range of energies up to $\gamma_L = 7$ (or $\gamma = 2$
for the identical nuclei). The obtained result has also a rather simple form~\eqref{final}. Certainly,
its asymptotic expression~\eqref{Fasym} completely coincides with Eq.~\eqref{final-2012}.

Using the obtained result we presented a detailed description for the NICA collider in Subsection 3.1.
It should be noted that all the cross sections above are given for the production of the $e^+ e^-$ pair
with the capture of the electron by the {\it second nucleus}. To obtain the total bound-free cross
section we must add the cross section with the capture of the electron by the {\it first nucleus} as well.
It results in doubling the numbers presented in Table~3 and Fig.~\ref{Fig5}.  In particular, the cross section of the BFPP in Bi$^{83+}$--Bi$^{83+}$ collisions at the NICA collider with the ion energy $5$ GeV/nucleon is equal to $60$ barn (taking into account the capture of an electron by both nuclei). This number is approximately one order of magnitude larger that the corresponding number for the hadronic interaction for which $\sigma_{\rm hadr}\approx 7$ barn.

It is useful to note that the capture to the excited atomic states adds about 10\% in the total cross sections.
%
%
\section*{Acknowledgments}

We are grateful to I.~Meshkov who attracted our attention to this problem and
explained us the details of the NICA collider. The stimulating discussions
with A.~Milstein, V.~Parkhomchuk and A.~Surzhykov are gratefully acknowledged.

%
%
%
%

\end{document}